# Robust Square Root Unscented Kalman filter of graph signals

Jinhui Hu, Haiquan Zhao, Senior *Member, IEEE,* and Yi Peng

*Abstract*—Considering the problem of nonlinear and non-gaussian filtering of the graph signal, in this paper, a robust square root unscented Kalman filter based on graph signal processing is proposed. The algorithm uses a graph topology to generate measurements and an unscented transformation is used to obtain the priori state estimates. In addition, in order to enhance the numerical stability of the unscented Kalman filter, the algorithm combines the double square root decomposition method to update the covariance matrix in the graph frequency domain. Furthermore, to handle the non-Gaussian noise problem in the state estimation process, an error augmentation model is constructed in the graph frequency domain by unifying the measurement error and state error, which utilizes the Laplace matrix of the graph to effectively reduce the cumulative error at each vertex. Then the general robust cost function is adopted as the optimal criterion to deal with the error, which has more parameter options so that effectively suppresses the problems of random outliers and abnormal measurement values in the state estimation process. Finally, the convergence of the error of the proposed algorithm is firstly verified theoretically, and then the robustness of the proposed algorithm is verified by experimental simulation.

*Index Terms*—unscented Kalman filter, graph signals processing, M-estimation, state estimation.

## 1. Introduction

Kalman filtering is a commonly used tool for state estimation [1]. The performance of the filter degrades when Kalman filtering implements state estimation on the state space model of a high dimensional graph signal [2][3]. Graph filters are used to efficiently analyze irregular as well as complex data, which effectively process high-dimensional graph signals by expanding the graph signals to the vertices of the graph and processing the information between the vertices in conjunction with the graph Laplacian matrix [4]-[8]. Therefore, using a combination of graph filters and Kalman filters help Kalman filtering to effectively deal with high dimensional graph signaling problems [2][3].

To solve the problem, unscented Kalman filtering based on graph signals (GSP-UKF) [3] as well as extended Kalman filtering based on graph signals (GSP-EKF) [2] algorithms have been developed. However, the development of GSP-KF

This work was supported in part by National Natural Science Foundation of China (Grant: 62171388, 61871461, 61571374). (Corresponding author: Haiquan Zhao).

The authors are with Key Laboratory of Magnetic Suspension Technology and Maglev Vehicle, Ministry of Education, School of Electrical Engineering, Southwest Jiaotong University, Chengdu 610031, China (e-mail: jhhu_swjtu@126.com; hqzhao_swjtu@126.com, pengyi1007@163.com).

has the same problem as the current development of robust KF, which also deals with the noise problem encountered in the state estimation process according to the Gaussian assumption, and the performance of the currently developed GSP-UKF and GSP-EKF algorithms degrades when faced with the non-Gaussian noise problem, and more accurate state estimation results cannot be realized. Therefore, it is necessary to introduce a robust approach in the development of GSP-KFs to effectively deal with the non-Gaussian noise problem encountered during state estimation. Various existing robust KF algorithms have demonstrated strong performance in various state estimation environments, such as the Gaussian MCCKF [9], MEEKF [10], and various robust algorithms developed by combining KF variants [11]-[15]. Due to the ability to be sensitive to great anomalies, M-estimation have shown powerful results in various fields [18][19][20]. M-estimation effectively eliminate the effect of large errors by assigning weights to the errors according to the magnitude of the errors [18]. Therefore, utilizing M-estimation to deal with the error can make the Kalman filtering algorithm avoiding be affected by the problem of impulsive noise, and noise from sudden changes in the measurements, thus making it more robust. A general robust loss function plays a role in various fields [13][21][22]. With different parameter choices, this cost function can model different M-estimated cost functions. Therefore, the use of this cost function will be adapted to more complex noise environments.

Also requiring attention is the development of nonlinear Kalman filtering algorithms. Commonly used nonlinear Kalman filtering algorithms include extended Kalman filtering (EKF) [2], unscented Kalman filtering (UKF) [1][3], and cubature Kalman filtering (CKF) [23], all of which have demonstrated high accuracy in nonlinear state estimation problems. However, when the sigma points are generated by using the UKF as well as the CKF algorithm, it is necessary to pay attention to the problem of decomposing the covariance matrix, and when encountering the complex noise environment or the UKF without unscented points is not selected properly, the covariance matrix will be non-positively determined, which cannot effectively ensure the stable operation of the algorithm. Therefore, a method based on double square root decomposition to update the error covariance matrix is proposed to address the above problems. Through this, algorithms such as SRCKF [23], SRUKF [24], etc. are developed, which perform well in the power system state estimation problem.

In this paper, a robust square root decomposition unscented Kalman filtering based on graph signal processing is proposed.

The algorithm adopts the method of graph filter to cope with the state estimation problem of high-dimensional systems, and also considers the non-Gaussian noise problem encountered during the state estimation process, adopts the general robust loss function for processing, and constructs an error augmentation model in the graph frequency space to make the error generated by the state process more sensitive. The numerical stability of the algorithm and its performance in dealing with different noise environments are further improved by using the double square root decomposition. Eventually, the robustness of the algorithm is validated through simulation. The main contributions of this paper are as follows

1) A robust square root unscented Kalman filtering algorithm of graph signal processing based on general robust loss function is proposed. It effectively handles the non-Gaussian noise problem in high-dimensional state space, and a double square root decomposition is used to improve the numerical stability of the algorithm.

2) By analyzing the error convergence, the stability of the proposed algorithm is determined theoretically.

3) The experimental results verify that the proposed algorithm is more robust with respect to the existing algorithms.

The rest of the paper is organized as follows, the model of the graph filter is presented in Section II, the derivation of the algorithm is given in Section III, an analysis of the stability of the algorithm is presented in Section IV, experimental results are given in Section V, and finally conclusions are given in Chapter VI.

## 2. Graph signal processing And General robust loss function

### 2.1. Graph signal processing

Consider an undirected connected weighted graph $\mathcal{G}(\mathcal{V}, \xi, \mathbf{A})$ with N vertices, where $\mathcal{V}$ and $\xi$ are the sets of vertices and edges respectively. $\mathbf{A}$ is a nonnegative weighted adjacency matrix, and $\mathrm{L} = \mathrm{diag}(\mathbf{A}) - \mathbf{A}$ represents the Laplace matrix of the graph, where $\mathrm{diag}(\mathbf{A})$ is the diagonal matrix of $\mathbf{A}$ whose (i, i)-th term is $a_i$. Since the Laplace matrix L is semi-positive definite, it is eigen valued decomposed [2][3][4]

$$\mathrm{L} = \mathbf{V} \Delta \mathbf{V}^T \tag{1}$$

where $\Delta$ is the matrix consisting of the eigenvalues of L, and $\mathbf{V}$ is L matrix consisting of the eigenvectors of b and satisfying $\mathbf{V}^T = \mathbf{V}^{-1}$. When the graph topology is determined, the corresponding graph Laplace matrix is also determined.

In this paper, the state quantities in the state estimation process are expanded onto the graph vertices, i.e., a scalar value is assigned to each vertex, and the dynamical equations used is

$$\mathbf{x}_i = \mathbf{f}\left(\mathbf{x}_{i-1}\right) + \mathbf{q}_i \tag{2}$$

where $\mathbf{x}_i \in \mathbb{R}^{N \times 1}$ represents the state quantity at $i$-th moment, $\mathbf{f}(\bullet)$ is the state transfer equation, $\mathbf{q}_i$ represents Gaussian

noise with 0 mean and the state noise covariance matrix is $\mathbf{Q}_i$. The measurement equation is

$$\mathbf{y}_i = \mathbf{h}\left(\mathbf{x}_i\right) + \mathbf{r}_i \tag{3}$$

where $\mathbf{y}_i \in \mathbb{R}^{N \times 1}$ represents the observation vector at moment i, $\mathbf{h}(\bullet)$ is the measurement equation, and $\mathbf{r}_i$ is Gaussian noise with zero mean independent of $\mathbf{q}_i$. The measurement noise covariance matrix is $\mathbf{R}_i$.

### 2.2. General robust loss function

Generalized robust loss functions effectively handle various types of non-Gaussian noise due to their diverse parameter choices and their sensitivity to error. Therefore, in this paper, we adopt this cost function as the optimization criterion of the algorithm. It takes the following form [21][22]

$$\varphi(c) = \frac{|\beta - 2|}{\beta} \left( \left( \frac{(c/\gamma)^2}{|\beta - 2|} + 1 \right)^{\beta/2} - 1 \right) \tag{4}$$

where $\beta$ is the shape factor, and $\gamma$ is a scale parameter. By changing the value of $\beta$, it will be converted to a different type of M-estimation cost function. For instance, the function becomes the Huber cost function when $\beta = 2$, as well as when $\beta = 0$, the cost function has the same form as Cauchy [22]. Therefore, it will be adapted to more complex noise environments compared to other M-estimation cost functions.

**Remark 1:** M-estimation can weight the error points according to the size of the error, thus effectively removing the interference of large outliers. The general robust loss function is a generalized form of the M-estimation cost function, and the different values of its parameters give it the form of various types of different M estimation cost functions. Therefore, compared with the general cost function, the generalized robust loss function has stronger robustness and can adapt to various complex noise environments [21][22].

## 3. Robust Square Root Unscented Kalman filter of graph signals based on General robust loss function

### 3.1. Robust Square Root Unscented Kalman filter of graph signals based on General robust loss function

In this section, the algorithmic steps are derived for a robust square root unscented kalman filter of graph signals based on General robust loss function (GSP-GR-SRUKF). For the state space equations in (2) and (3), the unscented transform [1] is used to realize the update of the priori state estimates as well as the priori covariance matrix.

1) Prediction: The UT transformation is used to generate $2N+1$ sigma points near the posteriori $\mathbf{x}_{i-1|i-1}$.

$$\chi_{i-1|i-1}^s = \begin{cases} \mathbf{x}_{i-1|i-1} & s = 0 \\ \mathbf{x}_{i-1|i-1} + \left(\sqrt{(N+\eta)}\Sigma_{i-1|i-1}\right)_s & s = 1, \dots, N \\ \mathbf{x}_{i-1|i-1} + \left(\sqrt{(N+\eta)}\Sigma_{i-1|i-1}\right)_s & s = N+1, \dots, 2N \end{cases} \tag{5}$$

where $\Sigma_{i-1|i-1} = \sqrt{\mathbf{P}_{i-1|i-1}}$ , $\mathbf{P}_{i-1|i-1}$ is the covariance matrix. $\eta$ is then the proportional correction parameter, which is used to adjust the position of the sigma point. An update to the a priori state estimate

$$\mathbf{x}_{i|i-1} = \sum_{s=0}^{2N} \omega_m^s \mathbf{f}\left(\chi_{i-1|i-1}^s\right) \quad (6)$$

The priori covariance matrix is then updated using the square root decomposition method

$$\Sigma_{i|i-1} = qr\left\{\left[\sqrt{\omega_c^1}\left(\chi_{1:2n,i|i-1} - \mathbf{x}_i^-\right)\right] \quad \sqrt{\mathbf{Q}_{i-1}}\right\} \quad (7)$$

$$\Sigma_{i|i-1} = cholupdate\left\{\Sigma_{i|i-1} \quad \chi_{0,i} - \mathbf{x}_i^- \quad \omega_c^0\right\} \quad (8)$$

where $\omega_m^o = \dfrac{\eta}{N+\eta}$ , $\omega_c^0 = \dfrac{\eta}{N+\eta} + (1-\alpha^2+\beta)$ , $\omega_m^s = \omega_c^s = \dfrac{1}{2(N+\eta)}, s = 1 \sim 2n$ . $qr\{\bullet\}$ refers to the qr decomposition, a command function in Matlab. $cholupdate\{\bullet\}$ is also a command function of Matlab for implementing Cholesky decomposition [23][24].

Remark 2: Here and in equations (12), (13), the covariance matrix is updated using square root decomposition, which will effectively help the algorithm to improve numerical stability [23][24].

2) Update: The matrix of update step is first handled by graph Fourier transform. UT transformation is used to generate 2N+1 sigma points in the neighborhood of the priori $\mathbf{x}_{i|i-1}$

$$\chi_{i|i-1}^s = \begin{cases} \mathbf{x}_{i|i-1} & s=0 \\ \mathbf{x}_{i|i-1} + \left(\sqrt{(N+\eta)}\Sigma_{i|i-1}^v\right)_s & s=1,\dots,N \\ \mathbf{x}_{i|i-1} + \left(\sqrt{(N+\eta)}\Sigma_{i|i-1}^v\right)_s & s=N+1,\dots,2N \end{cases} \quad (9)$$

Then the measurement estimates as well as the mutual covariance matrix is computed by graph Fourier transform

$$\overset{v}{\mathbf{y}}_{i|i-1} = \sum_{s=0}^{2N} \omega_m^s \mathbf{V}^T \mathbf{h}\left(\chi_{i|i-1}^s\right) \quad (10)$$

$$\mathbf{P}_{xy,i|i-1}^v = \sum_{s=0}^{2N} \omega_m^s \left(\mathbf{V}^T\left(\chi_{i|i-1}^s - \mathbf{x}_{i|i-1}\right)\right)\left(\mathbf{V}^T\left(\mathbf{h}\left(\chi_{i|i-1}^s\right) - \mathbf{y}_{i|i-1}\right)\right)^T \quad (11)$$

$$\Omega_{i|i-1}^v = qr\left\{\left[\sqrt{\omega_c^1}\mathbf{V}^T\left(\mathbf{y}_{1:2n,i|i-1} - \mathbf{y}_i\right)\right] \quad \mathbf{V}^T\sqrt{\mathbf{R}_{i-1}}\right\} \quad (12)$$

$$\Omega_{i|i-1}^v = cholupdate\left\{\Omega_{i|i-1}^v \quad \chi_{0,i} - \mathbf{x}_i^- \quad \omega_c^0\right\} \quad (13)$$

where $\Omega_{i|i-1}^v = \sqrt{\mathbf{P}_{yy,i|i-1}^v}$ , $\mathbf{P}_{yy,i|i-1}^v$ is the measurement error covariance matrix.

The covariance matrix, state estimates, and measurements are then transformed by graph Fourier transform

$$\mathbf{P}_{i|i-1}^v = \mathbf{V}^T \mathbf{P}_{i|i-1} \mathbf{V} \quad (14)$$

$$\overset{v}{\mathbf{x}}_{i|i-1} = \mathbf{V}^T \mathbf{x}_{i|i-1} \quad (15)$$

$$\mathbf{y}_i^v = \mathbf{V}^T \mathbf{y}_i \quad (16)$$

$$\mathbf{Q}_i^v = \mathbf{V}^T \mathbf{Q}_i \mathbf{V} \quad (17)$$

$$\mathbf{R}_i^v = \mathbf{V}^T \mathbf{R}_i \mathbf{V} \quad (18)$$

Next, an error augmentation model is constructed in the graph frequency domain to unify the state and measurement errors.

$$\begin{bmatrix} \overset{v}{\mathbf{x}}_{i|i-1} \\ \overset{v}{\mathbf{z}}_i \end{bmatrix} = \begin{bmatrix} \mathbf{x}_i^v \\ \mathbf{H}_i^v \mathbf{x}_i^v \end{bmatrix} + \begin{bmatrix} -(\mathbf{x}_i^v - \overset{v}{\mathbf{x}}_{i|i-1}) \\ \mathbf{r}_i^v \end{bmatrix} \quad (19)$$

where $\mathbf{H}_i^v = \left(\left(\mathbf{P}_{i|i-1}^v\right)^{-1} \mathbf{P}_{xy,i|i-1}^v\right)^T$ .

Remark 3: The various signals can be converted to signals in the graph frequency domain by taking the graph Fourier transform of each matrix. Graph Fourier transforming the error information allows each vertex information to be updated independently to reduce the cumulative error [2] [3]. Therefore, the graph Fourier transformed error information is used to build the error generalization model, which will help the algorithm to achieve independent processing of each vertex error in the subsequent error processing to reduce the cumulative error of state estimation.

The covariance matrix of $\begin{bmatrix} -(\mathbf{x}_i^v - \overset{v}{\mathbf{x}}_{i|i-1}) \\ \mathbf{r}_i^v \end{bmatrix}$ are

$$\mathbf{B}_i^v = E\left[\begin{bmatrix} -(\mathbf{x}_i^v - \overset{v}{\mathbf{x}}_{i|i-1}) \\ \mathbf{r}_i^v \end{bmatrix}\begin{bmatrix} -(\mathbf{x}_i^v - \overset{v}{\mathbf{x}}_{i|i-1}) \\ \mathbf{r}_i^v \end{bmatrix}^T\right] = \begin{bmatrix} \mathbf{P}_{i|i-1}^v & 0 \\ 0 & \mathbf{R}_i^v \end{bmatrix} \quad (20)$$

By Cholesky decomposition, one obtains

$$\mathbf{B}_i^v = \Upsilon_i^v\left(\Upsilon_i^v\right)^T = \begin{bmatrix} \Psi_{i|i-1}^v\left(\Psi_{i|i-1}^v\right)^T & 0 \\ 0 & \Phi_i^v\left(\Phi_i^v\right)^T \end{bmatrix} \quad (21)$$

where $\Upsilon_i^v$, $\Psi_{i|i-1}^v$ and $\Phi_i^v$ are the Chole sky decomposition of $\mathbf{B}_i^v$, $\mathbf{P}_{i|i-1}^v$ and $\mathbf{R}_i^v$ respectively. Then left-multiplying $\left(\Upsilon_i^v\right)^{-1}$ by (19) yields

$$\mathbf{d}_i^v = \Gamma_i^v \mathbf{x}_i^v + e_i^v \quad (22)$$

where

$$\mathbf{d}_i^v = \left(\Upsilon_i^v\right)^{-1}\begin{bmatrix} \overset{v}{\mathbf{x}}_{i|i-1} \\ \overset{v}{\mathbf{z}}_i \end{bmatrix} \in \mathbb{R}^{2N \times 1} \quad (23)$$

$$\Gamma_i^v = \left(\Upsilon_i^v\right)^{-1}\begin{bmatrix} \mathbf{I} \\ \mathbf{H}_i^v \end{bmatrix} \in \mathbb{R}^{2N \times 1} \quad (24)$$

$$e_i^v = \left(\Upsilon_i^v\right)^{-1}\begin{bmatrix} -(\mathbf{x}_i^v - \overset{v}{\mathbf{x}}_{i|i-1}) \\ \mathbf{r}_i^v \end{bmatrix} \in \mathbb{R}^{2N \times 1} \quad (25)$$

Since $E\left[e_i^v\left(e_i^v\right)^T\right] = I$ , therefore $e_i^v$ is white.

Then the optimal value of the state is obtained by

$$\overset{v}{\mathbf{x}}_{i|i} = \arg\min g\left(\mathbf{x}_i^v\right) \quad (26)$$

where the cost function $g\left(\mathbf{x}_i^v\right)$ is induced by natural regression and is given by the following equation

---

**Algorithm 1: GSP-GR-SRUKF**

**1: Input:** $\mathbf{f}(\bullet)$, $\mathbf{h}(\bullet)$, $\mathbf{Q}_i$, $\mathbf{R}_i$, $\xi$, $\rho(\bullet)$, $\mathbf{V}$.

**2: Output:** $\mathbf{x}_{i|i}$ for $i = 1, 2, \ldots, N$

**3: Intitialization:** Setting initial filter state values $\mathbf{x}_{0|0}$ and initial square root of the covariance matrix values $\Sigma_{0|0}$.

**4: for** $i = 1, 2, \ldots, N$ **do**

> The priori state estimates $\mathbf{x}_{i|i-1}$ and square root of the priori covariance matrix $\Sigma_{0|0}$ are computed by (5)-(13).
>
> Using (14)-(18) to transform the covariance matrix from the algorithm as well as the state values into the graph frequency domain.
>
> Let $\left(\mathbf{x}_{i|i}^{v}\right)^{0} = \mathbf{x}_{i|i-1}^{v}$, and compute $\left(\mathbf{x}_{i|i}^{v}\right)^{1}$ by (36).
>
> **while** $\dfrac{\left\| \left(\mathbf{x}_{i|i}^{v}\right)^{j+1} - \left(\mathbf{x}_{i|i}^{v}\right)^{j} \right\|}{\left\| \left(\mathbf{x}_{i|i}^{v}\right)^{j} \right\|} > \xi$ **do**
>
> Compute the $\left(\mathbf{x}_{i|i}^{v}\right)^{j+1}$ via (36).
>
> **end while**
>
> Compute actual values $\mathbf{x}_{i|i} = \mathbf{V}^{T} \mathbf{x}_{i|i}^{v}$.
>
> Updating square root of the covariance matrix by (40).

**end for**

---

$$g\left(\mathbf{x}_{i}^{v}\right) = \sum_{k=1}^{2N} \varphi\left(e_{i,k}^{v}\right) \tag{27}$$

where $\varphi(\bullet)$ is given by (4), $e_{i,k}^{v}$ represents the $k$-th element of $e_{i}^{v}$. By finding the gradient of (26), the value of state $\mathbf{x}_{i|i}^{v}$ is obtained by

$$\mathbf{x}_{i|i}^{v} = \sum_{k=1}^{2N} \frac{\partial \varphi\left(e_{i,k}^{v}\right)}{\partial e_{i,k}^{v}} \frac{\partial e_{i,k}^{v}}{\partial \mathbf{x}_{i}^{v}} \tag{28}$$

Letting

$$\rho\left(e_{i,k}^{v}\right) = \frac{\partial \varphi\left(e_{i,k}^{v}\right)}{\partial e_{i,k}^{v}} \frac{1}{e_{i,k}^{v}} \tag{29}$$

**Remark 4:** The algorithm uses a general robust loss function as the optimality criterion for handling errors. When the cost function $\rho\left(e_{i,k}^{v}\right) = 1$ is chosen here, the algorithm is not robust and degenerates to the square root UKF of graph signal processing (GSP-SRUKF).

In order to proceed, it is necessary to consider the following matrix.

$$\Xi_{\mathbf{x},i}^{v} = diag\left[ \rho\left(e_{i,1}^{v}\right), \rho\left(e_{i,2}^{v}\right), \ldots, \rho\left(e_{i,N}^{v}\right) \right] \tag{30}$$

$$\Xi_{\mathbf{y},i}^{v} = diag\left[ \rho\left(e_{i,N+1}^{v}\right), \rho\left(e_{i,N+2}^{v}\right), \ldots, \rho\left(e_{i,2N}^{v}\right) \right] \tag{31}$$

with

$$\Xi_{i}^{v} = \begin{bmatrix} \Xi_{\mathbf{x},i}^{v} & 0 \\ 0 & \Xi_{\mathbf{y},i}^{v} \end{bmatrix} \tag{32}$$

Since $\mathbf{d}_{i}^{v}$ and $\mathbf{x}_{i}^{v}$ are independent of each other, there are

$$\frac{\partial e_{i,k}^{v}}{\partial \mathbf{x}_{i}^{v}} = -\Gamma_{i}^{v} \tag{33}$$

$$\sum_{k=1}^{2N} \frac{\partial \varphi\left(e_{i,k}^{v}\right)}{\partial e_{i,k}^{v}} \frac{\partial e_{i,k}^{v}}{\partial \mathbf{x}_{i}^{v}} = \Gamma_{i}^{v} \Xi_{i}^{v}\left(\Gamma_{i}^{v} \mathbf{x}_{i}^{v} - \mathbf{d}_{i}^{v}\right) \tag{34}$$

Making the gradient of (26) equal to 0, the optimal value of the state is obtained by

$$\mathbf{x}_{i|i}^{v} = \left(\Gamma_{i}^{v} \Xi_{i}^{v} \Gamma_{i}^{v}\right)^{-1} \Gamma_{i}^{v} \Xi_{i}^{v} \mathbf{d}_{i}^{v} \tag{35}$$

According to the matrix inversion lemma, we have

$$\mathbf{x}_{i|i}^{v} = \mathbf{x}_{i|i-1}^{v} + \mathbf{K}_{i}^{v}\left(\mathbf{y}_{i}^{v} - \mathbf{y}_{i|i-1}^{v}\right) \tag{36}$$

where

$$\mathbf{K}_{i}^{v} = \overline{\mathbf{P}}_{i|i-1}^{v} \mathbf{H}_{i}^{v}\left(\mathbf{H}_{i}^{v} \overline{\mathbf{P}}_{i|i-1}^{v}\left(\mathbf{H}_{i}^{v}\right)^{T} + \overline{\mathbf{R}}_{i}^{v}\right) \tag{37}$$

with

$$\overline{\mathbf{P}}_{i|i-1}^{v} = \Upsilon_{i}^{v}\left(\Xi_{\mathbf{x},i}^{v}\right)^{-1}\left(\Upsilon_{i}^{v}\right)^{T} \tag{38}$$

$$\overline{\mathbf{R}}_{i}^{v} = \Upsilon_{i}^{v}\left(\Xi_{\mathbf{y},i}^{v}\right)^{-1}\left(\Upsilon_{i}^{v}\right)^{T} \tag{39}$$

Finally, updating the covariance matrix using the square root decomposition

$$\Sigma_{i|i}^{v} = qr\left(\left[\left(\mathbf{I} - \mathbf{K}_{i}^{v} \mathbf{H}_{i}^{v}\right)\Sigma_{i|i-1}^{v}, \ \mathbf{K}_{i}^{v} \overline{\mathbf{R}}_{i}^{v}\right]\right) \tag{40}$$

Furthermore, in order to reduce the computational burden and optimize $\mathbf{K}_{i}^{v}$, one can do so by solving for the partial derivatives of $trace\left(\mathbf{P}_{i|i-1}^{v}\right)$ with respect to $\mathbf{K}_{i}^{v}$ and making the partial derivatives 0. We have [2]

$$\mathbf{K}_{i}^{v} = diag\left(\overline{\mathbf{P}}_{i|i-1}^{v} \mathbf{H}_{i}^{v}\right) diag\left(\left(\mathbf{H}_{i}^{v} \overline{\mathbf{P}}_{i|i-1}^{v}\left(\mathbf{H}_{i}^{v}\right)^{T} + \overline{\mathbf{R}}_{i}^{v}\right)^{-1}\right) \tag{41}$$

The above is the whole process of deriving the GSP-GR-SRUKF algorithm, followed by a summary of the algorithm as in Algorithm 1.

### 3.2. Stability analysis

To verify the error convergence of the proposed algorithm, first defining an error factor in the graph frequency domain

$$\xi_{i} = \mathbf{x}_{i}^{v} - \mathbf{x}_{i}^{v} \tag{42}$$

Combining (2), (3), (36) and (41), (42) is rewritten as

$$\xi_{i} = \left(\mathbf{I} - \mathbf{K}_{i}^{v} \mathbf{H}_{i}^{v}\right)\mathbf{F}_{i-1}^{v} \xi_{i-1} + \left(\mathbf{I} - \mathbf{K}_{i}^{v} \mathbf{H}_{i}^{v}\right)\mathbf{q}_{i}^{v} - \mathbf{K}_{i}^{v} \mathbf{r}_{i}^{v} \tag{43}$$

where $\mathbf{F}_{i-1}^{v} = \mathbf{V}^{T}\left(\dfrac{\partial \mathbf{f}(\mathbf{x}_{i-1})}{\partial \mathbf{x}_{i-1}}\right)\mathbf{V}$, $\mathbf{q}_{i}^{v} = \mathbf{V}^{T} \mathbf{q}_{i}$, $\mathbf{r}_{i}^{v} = \mathbf{V}^{T} \mathbf{r}_{i}$.

Considering that $\mathbf{q}_{i}$ and $\mathbf{r}_{i}$ are both zero-mean Gaussian white noise, the expectation of $\xi_{i}$ is

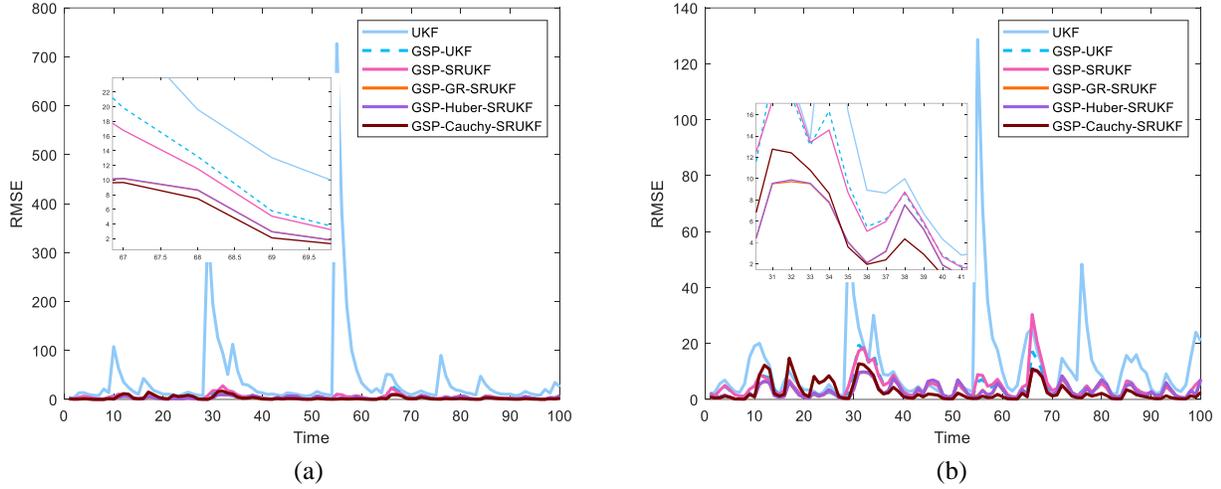

(a)

(b)

Fig. 1 (a) RMSE of different algorithm in Gaussian noise of 1 variance
(b) RMSE of different algorithm in Gaussian noise of 100 variance

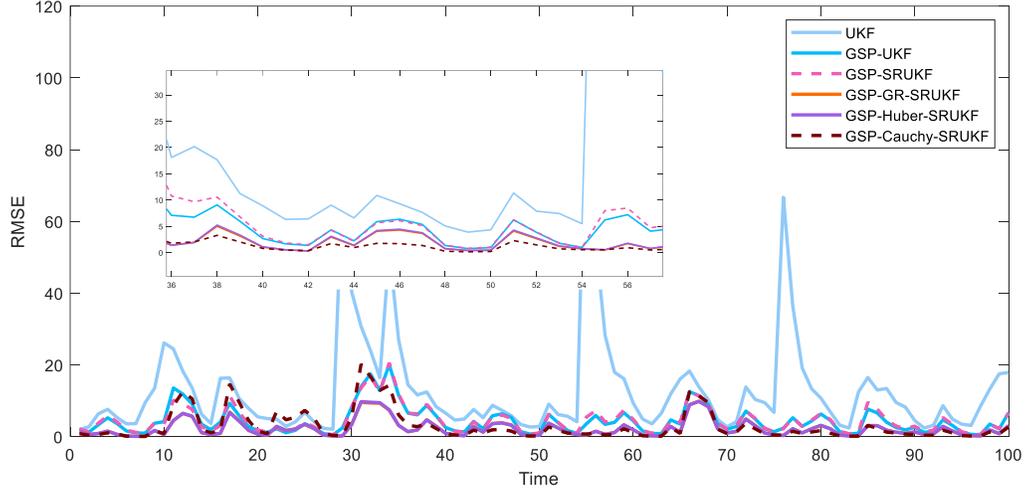

Fig. 2. RMSE of different algorithm in noise $r_{1,i}$

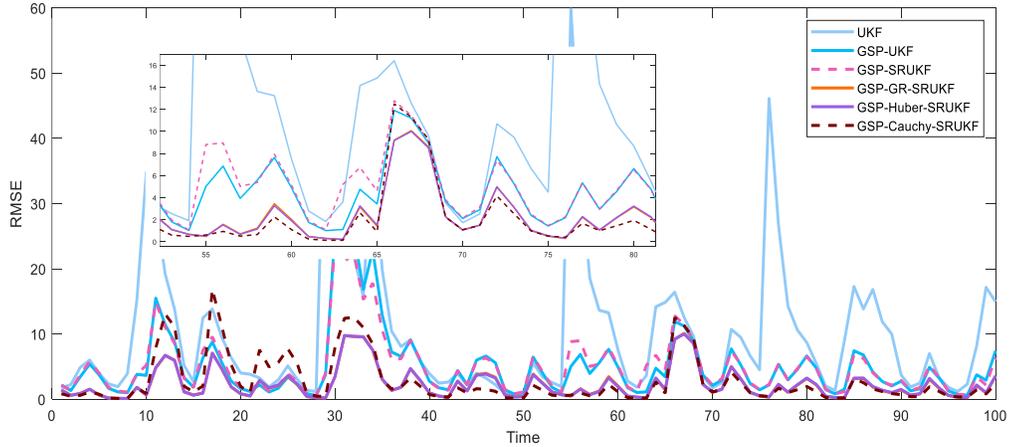

Fig. 3. RMSE of different algorithm in noise $r_{3,i}$

$$E[\xi_i] = \left(\mathbf{I} - \mathbf{K}_i^y \mathbf{H}_i^y\right) \mathbf{F}_{i-1}^y E[\xi_{i-1}] \qquad (44)$$

Since the state update equations as well as the measurement equations in the state space equations considered in this paper are time-invariant, and the adopted graph topology is determined in the algorithm, $\mathbf{H}_i^v$ and $\mathbf{F}_i^v$ are stable. Next, according to (26)-(41), we obtain that $\mathbf{K}_i^v$ is also stable, and thus, (44) shows that the algorithm's mean-error behavior is convergent.

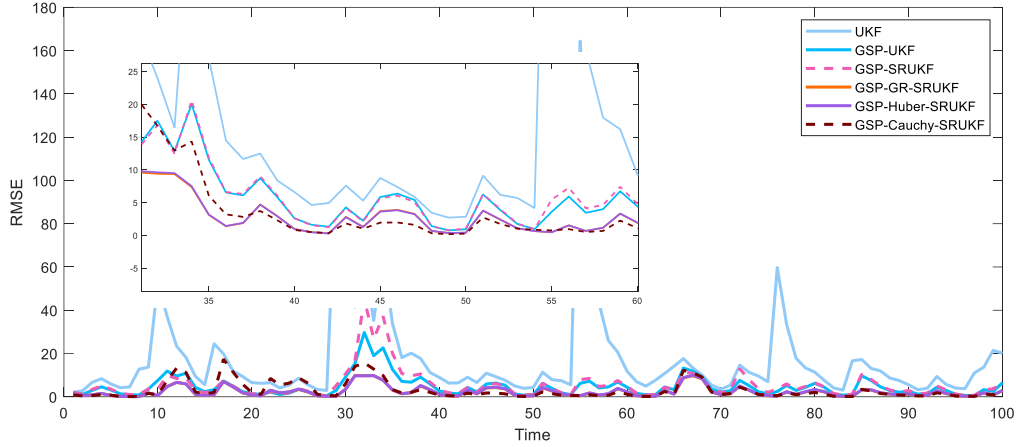

Fig. 4. RMSE of different algorithm in noise $r_{4,i}$

Furthermore, considering the mean square error behavior, the error covariance matrix is written as

$$E\left[\xi_i \xi_i^T\right] = \left(\mathbf{I} - \mathbf{K}_i^v \mathbf{H}_i^v\right) \mathbf{F}_{i-1}^v E\left[\xi_{i-1} \xi_{i-1}^T\right] \left(\mathbf{F}_{i-1}^v\right)^T \left(\mathbf{I} - \mathbf{K}_i^v \mathbf{H}_i^v\right)^T$$
$$+ \left(\mathbf{I} - \mathbf{K}_i^v \mathbf{H}_i^v\right) \mathbf{Q}_i^v \left(\mathbf{I} - \mathbf{K}_i^v \mathbf{H}_i^v\right)^T - \mathbf{K}_i^v \mathbf{R}_i^v \left(\mathbf{K}_i^v\right)^T \quad (45)$$

Letting

$$\delta_i = E\left[\xi_i \xi_i^T\right] \quad (46)$$

$$\mathbf{A}_i = \left(\mathbf{I} - \mathbf{K}_i^v \mathbf{H}_i^v\right) \mathbf{F}_{i-1}^v \quad (47)$$

$$\mathbf{B}_i = \left(\mathbf{I} - \mathbf{K}_i^v \mathbf{H}_i^v\right) \mathbf{Q}_i^v \left(\mathbf{I} - \mathbf{K}_i^v \mathbf{H}_i^v\right)^T - \mathbf{K}_i^v \mathbf{R}_i^v \left(\mathbf{K}_i^v\right)^T \quad (48)$$

Therefore, (45) can be written as

$$\delta_i = \mathbf{A}_i \delta_{i-1} \mathbf{A}_i^T + \mathbf{B}_i \quad (49)$$

Since $\mathbf{H}_i^v$, $\mathbf{F}_{i-1}^v$, $\mathbf{Q}_i^v$, and $\mathbf{R}_i^v$ are time-invariant and the Kalman gain $\mathbf{K}_i^v$ is stable, $\mathbf{A}_i$ and $\mathbf{B}_i$ are stable and $\delta_i$ is convergent. We make $\lim_{i \to \infty} \delta_i = \delta$ , $\lim_{i \to \infty} \mathbf{A}_i = \mathbf{A}$ , $\lim_{i \to \infty} \mathbf{B}_i = \mathbf{B}$ . Therefore, (49) can be reduced to the real discrete-time Liapunov equation

$$\delta = \mathbf{A} \delta \mathbf{A}^T + \mathbf{B} \quad (50)$$

with

$$vec(\delta) = \left(\mathbf{I} - \mathbf{A} \otimes \mathbf{A}^T\right)^{-1} + vec(\mathbf{B}) \quad (51)$$

where I represent an identity matrix and $\otimes$ denotes the Kronecker product.

So far, based on the mean error behavior and the mean square error behavior of the algorithm, determining the convergence of the algorithm's error and we will conclude that the algorithm is theoretically error convergent.

## 4. Numerical Results

In this section, the robustness of the proposed GSP-GR-SRUKF is tested. The robustness of our algorithm is verified by comparing it with UKF [3] and GSP-UKF [3] algorithms in various noisy environments. In order to ensure that the experiments have higher statistical properties, $M$=1000 independent Monte Carlo experiments is conducted.

Consider the following nonlinear system [3]

$$\mathbf{x}_{i,k} = \frac{1}{2} \mathbf{x}_{i-1,k} + \frac{25 \mathbf{x}_{i-1,k}}{1 + \mathbf{x}_{i-1,k}^2} + 8 \cos\left(1.2(i-1)\right) + \mathbf{q}_{i-1,k} \quad (52)$$

$$\mathbf{y}_{i,k} = \mathbf{x}_{i,k} + \phi \sin\left(\mathbf{x}_{i,k}\right) + \frac{\phi}{\mathbf{x}_{i,k} + \mathbf{x}_{i,k}^2} + \mathbf{r}_{i,k} \quad (53)$$

where $\phi$ is a scalar. The initial state is set to $\mathbf{x}_{0,k} = 0.5k$ , where $k = 1,2,\ldots,N$ , $N$ is the dimension of the state. The initial state estimate is set to $\mathbf{x}_0 = \mathbf{x}_0 + \sigma$ . where $\sigma$ is Gaussian noise with 0 mean and 4 variance. In this section, Gaussian noise with a variance of 0.01 is added as state noise. The corresponding state noise covariance matrix is set as $\mathbf{Q}_i = 0.01 \mathbf{I}_N$ . A graph topology with vertices $N$=10 is generated to provide the graph for the algorithm.

The experimental metrics use the mean square error, which is given by

$$\text{RMSE}(i) = \sqrt{\frac{1}{N} \frac{1}{M} \sum_{k=1}^{M} \left(\left\|\mathbf{x}_i^k - \mathbf{x}_i^k\right\|\right)^2} \quad (54)$$

where $\mathbf{x}_i^k, \mathbf{x}_i^k$ are surrogates for the true and predicted states of the $k$-th Monte Carlo experiment at $i$-th moment, respectively. $N$ is the number of dimensions representing the state. Then further define an average RMSE (ARMSE)

$$\text{ARMSE} = \frac{1}{D} \sum_{i=1}^{N} \text{RMSE}(i) \quad (55)$$

where $D$ is the total time step.

Considering that the cost function in (29) plays a large role in the performance of an algorithm, different choices of cost function may lead to different performance of the algorithm. Comparing the general form of the two M-estimation cost functions, the same tests were performed using Huber and Cauchy functions to obtain two algorithms GSP-Huber-SRUKF and GSP-Cauchy-SRUKF. The two cost functions take the following form

1) Huber cost function

$$\rho(c) = \begin{cases} \dfrac{c^2}{2}, & |c| < \sigma \\ \sigma|c| - \dfrac{\sigma^2}{2} & |c| \geq \sigma \end{cases} \quad (56)$$

2) Cauchy cost function

$$\rho(c) = \dfrac{\sigma^2}{2}\log\left(1 + \dfrac{c^2}{\sigma}\right) \quad (57)$$

*Case A Gaussian noise*

The efficacy of the GSP-GR-SRUKF is initially evaluated in the presence of Gaussian noise in two distinct Gaussian noise environments, characterized by a variance of 1 and a variance of 100, respectively. Fig. 1 show the RMSE of the different algorithms, respectively. From the figure, it can be concluded that all three algorithms under the framework of the algorithm proposed in this paper has the same performance with GSP-UKF when facing Gaussian noise environment. And it will show better performance in the face of Gaussian noise with larger variance. This is due to the fact that the proposed algorithms in this paper construct the error augmentation and generalization model in the graph frequency domain, which incorporates the nature of the graph Laplace matrix, and reduce the cumulative error in the process of state updating. And the algorithm proposed in this paper uses square root decomposition to update the covariance matrix, which can effectively deal with Gaussian noise in the case of large variance. Therefore, the proposed algorithm obtains more accurate state estimation results.

*Case B Mixture Gaussian noise*

In order to further verify the robustness of GSP-GR-SRUKF to cope with the environment of mixture-Gaussian noise, two Gaussian noises with different variance sizes is used for mixing

$$\mathbf{r}_{1,i} \sim 0.99 \times N(0,10) + 0.01 \times N(0,10000) \quad (58)$$

where $N(0,10)$ represents Gaussian noise with a mean of 0 and a variance of 10. The RMSEs of different algorithms in this noisy environment are demonstrated in Fig. 2, and it can be concluded that all the three algorithms obtained in this paper using the M-estimation cost function as the optimality criterion have good robustness. Therefore, it can be concluded that the M-estimation approach can also achieve good results in dealing with non-Gaussian noise in the graph frequency domain.

A mixture of Gaussian noise with different mean-variance sizes is further used to produce a non-Gaussian noise with a more complex distribution.

$$\mathbf{r}_{2,i} \sim 0.99 \times N(-0.1,10) + 0.01 \times N(0.1,10000) \quad (59)$$

The ARMSEs of different algorithms are compared in Table II, and it will be concluded that the algorithm proposed in this paper copes well with this complex noise environment and maintain good robustness.

*Case C Asymmetric Noise*

In this subsection, Gaussian noise is utilized for mixing to obtain an asymmetric noise environment. The asymmetric noise model is given by

$$\mathbf{r}_{3,i} \sim 0.8 \times N(0,1) + 0.1 \times N(1,1000) + 0.1 \times N(-1,1000) \quad (60)$$

In this environment characterized by high levels of noise, the algorithm was tested to ascertain its resilience to non-Gaussian noise. The RMSEs of different algorithms in this noise environment are shown in Fig. 3, and all the three cost functions in the framework of the algorithm proposed in this paper show good robustness.

*Case D Non-Gaussian Noise*

To further verify the robustness of the proposed algorithm, two non-Gaussian noise distributions is used for further

TABLE I
ALGORITHM PARAMETER SETTING

| Algorithm | kernel width in nonlinear system |
|---|---|
| GSP-Cauchy-SRUKF | $\sigma = 1.1$ |
| GSP-Huber-SRUKF | $\sigma = 1.1$ |
| GSP-GR-SRUKF | $\beta = -1, \gamma = 1.1$ |

TABLE II
ARMSE UNDER NOISE $\mathbf{r}_{2,i}$ OF DIFFERENT ALGORITHM

| Algorithm | ARMSE |
|---|---|
| UKF | 14.038232 |
| GSP-UKF | 4.378922 |
| GSP-SRUKF | 4.403562 |
| GSP-Cauchy-SRUKF | **2.304637** |
| GSP-Huber-SRUKF | 2.558400 |
| GSP-GR-SRUKF | 2.341118 |

TABLE III
ARMSE UNDER NOISE $\mathbf{r}_{5,i}$ OF DIFFERENT ALGORITHM

| Algorithm | ARMSE |
|---|---|
| UKF | 29.169175 |
| GSP-UKF | 5.731030 |
| GSP-SRUKF | 6.253395 |
| GSP-Cauchy-SRUKF | 2.211721 |
| GSP-Huber-SRUKF | 2.559840 |
| GSP-GR-SRUKF | **2.202196** |

experiments. The first one is a stabilized distribution noise

$$\xi(k) = \exp\left\{ j\omega k - \delta|k|^\alpha \left[ 1 + j\beta sign(k) L(k,\alpha) \right] \right\} \quad (61)$$

where

$$L(k,\alpha) = \begin{cases} \tan\left(\dfrac{\alpha\pi}{2}\right), & \alpha \neq 1. \\ \dfrac{2}{\pi}\log|k|, & \alpha = 1. \end{cases} \quad (62)$$

where $\alpha$ represents the eigen factor, $\beta$ denotes the symmetry parameter, $\delta$ represents the dispersion parameter, and $\omega$ represents the position parameter. Define that the noise obeying $\alpha$ stable distribution can be expressed as

$$\mathbf{r}_i \sim \lambda(\alpha, \beta, \delta, \omega) \tag{63}$$

In this subsection, the following parameters are set for testing

$$\mathbf{r}_{s,i} \sim \lambda(1.2, 1, 0, 1) \tag{64}$$

The RMSE of different algorithms in this noisy environment is shown in Fig. 4, and it will be seen that the three robust algorithms proposed in this paper handle this noise well with a small RMSE.

Second, the Rayleigh noise distribution model is

$$r(k) = \frac{k}{\tau^2} \exp\left(-\frac{k^2}{2\tau^2}\right) \tag{65}$$

The Rayleigh noise distribution is defined as $r_i \sim S(\tau)$. The robustness of the algorithm is further verified using the noise model $r_{s,i} \sim S(3)$. The ARMSE of different algorithms is shown in Table III, and it can be seen that the algorithm proposed in this paper still has the best performance.

## 5. Conclusion

In this paper, a new robust Kalman filtering algorithm is developed in the graph frequency domain. The algorithm constructs an error augmentation model in the graph frequency domain with the objective of unifying the state error and measurement error. This approach effectively reduces the accumulation of cumulative error during the iteration process of the algorithm and improves the accuracy of the algorithm. And the general robust loss function is adopted as the optimal criterion for the algorithm error, which effectively removes the influence of large measurements in the state update process and makes the algorithm robust. In addition, by using square root decomposition to update covariance matrix, the numerical stability of the unscented Kalman filtering algorithm is improved in the graph frequency domain. Furthermore, this paper analyzes the error convergence of the proposed algorithm. Finally, the robustness of the proposed algorithm is verified in simulation experiments.